\documentclass{ws-mpla}
\usepackage[super]{cite}
\usepackage{bm}
\usepackage{mathrsfs}
\usepackage{bm}
\usepackage{multirow}
\usepackage{textcomp}

\begin{document}
	
	\markboth{H-F Ding, X-H Zhai}
	{Examining the weak cosmic censorship conjecture by gedanken experiments for an EMDA black hole}
	
	\catchline{}{}{}{}{}
	%
	%%%%%%%%%%%%%%%%%%%%%%%%%%%%%%%%%%%%%%%%%%%%%%%%%%%%%%%%%%%%%%%%%%%%

	\title{Examining the weak cosmic censorship conjecture by gedanken experiments for an Einstein-Maxwell-Dilaton-Axion black hole}
	
	\author{Hai-Feng Ding}
	
	\address{Division of Mathematical and Theoretical Physics, Shanghai Normal University, \\ 100 Guilin Road, Shanghai 200234, China \\ haifeng1116@qq.com}

	\author{Xiang-Hua Zhai}
	
	\address{Division of Mathematical and Theoretical Physics, Shanghai Normal University, \\ 100 Guilin Road, Shanghai 200234, China \\ zhaixh@shnu.edu.cn}
	
	\maketitle
	
	\begin{history}
		\received{Day Month Year}
		\revised{Day Month Year}
	\end{history}
	
	\begin{abstract}
		In the framework of the new version of the gedanken experiments proposed by Sorce and Wald, we investigate the weak cosmic censorship conjecture (WCCC) for an Einstein-Maxwell-Dilaton-Axion (EMDA) black hole. Our result shows that no violations of WCCC can occur with the increase of the background solution parameters for this near-extremal EMDA black hole when the second order correction of the perturbations is taken into account. Namely, the near-extremal EMDA black hole cannot be over-charged or over-spun.
	\end{abstract}
	
	
	\ccode{PACS numbers:}
	
	%\tableofcontents
	
	\section{Introduction}
	
	The weak cosmic censorship conjecture (WCCC), as one of the most important questions in classical gravitational theories, was proposed by Penrose \cite{Penrose:1969pc} to ensure the causality and the predictability of classical general relativity. It states that all singularities arising from gravitational collapse must be hidden behind event horizons of black holes and cannot be seen by distant observers.
	
	Several studies have been done to test the validity of WCCC \cite{Wald:1997wa}, and WCCC also attracted a lot of interests in various theories and by diverse methods \cite{Hod:2002pm,Jacobson:2009kt,Saa:2011wq,Jacobson:2010iu,Matsas:2007bj,Gao:2012ca,Li:2013sea,Duztas:2013wua,Duztas:2013gza,Toth:2015cda,Gwak:2015fsa,Gwak:2015sua,Revelar:2017sem,Liang:2018wzd,Gwak:2019rcz}, for a recent review and progresses see for instance Ref.~\citen{Ong:2020xwv} and references therein. Among these methods, one possible way of violating the WCCC is to over-charge or over-spin a black hole by adding the charged or spinning matter, so that the event horizon might be destroyed and the naked singularity would appear. In 1974, Wald proposed a gedanken experiment to examine the WCCC for an extremal Kerr-Newman black hole \cite{Wald:1974wa1}. It was shown that no violations of WCCC can occur by throwing particle matter into an extremal Kerr-Newman black hole. But in 1999, Hubeny proposed that violations of WCCC might still occur by over-charging a near-extremal charged black hole \cite{Hubeny:1998ga}. Recently, Sorce and Wald \cite{Sorce:2017dst} suggested that the analysis of Hubeny’s experiment is insufficient at the linear order, so that the second order correction of the perturbation must be taken into account to check the WCCC, and a new version of the gedanken experiments has been proposed.
	
	In the new version of the gedanken experiments \cite{Sorce:2017dst}, the detailed physical process of falling matter in Kerr-Newman spacetime has been analyzed based on the Iyer-Wald formalism \cite{Iyer:1994ys}. After the null energy condition of the falling matter was taken into account, they derived the first order and second order inequalities relating energy, angular momentum and electric charge of the black hole. Importantly, the second order inequality of mass automatically takes all effects on energy into account, including self-force and finite-size effects, and it is valid not only for particle-like matter but also for general matter entering a Kerr-Newman black hole. They showed that the near-extremal Kerr-Newman black hole cannot be over-charged or over-spun when the second order correction of the perturbation was taken into account.
	
	So far, there are no general procedures to prove the validity of WCCC for arbitrary black holes in any gravitational theories. Hence, its validity has to be demonstrated for black holes case by case. Most recently, by using the new version of the gedanken experiments, the WCCC has been tested for several stationary black hole solutions in Refs.~ \citen{An:2017phb,Ge:2017vun,Chen:2019nhv,Jiang:2019ige,Jiang:2019vww,Wang:2019bml,Jiang:2019soz,He:2019mqy,Jiang:2020btc,Jiang:2020mws,Wang:2020vpn,Shaymatov:2019del,Shaymatov:2020wtj}, and the results show that the WCCC is valid under the second order approximation. Usually, in the new version of the gedanken experiments, the background conserved charges include the mass, angular momentum and electric charge. It is worth further study whether the increase of the background solution parameters will affect the validity of WCCC. The Einstein-Maxwell-Dilaton-Axion (EMDA) black hole \cite{Garcia:1995qz} with stationary and axial symmetries is equipped with six continuous free parameters and two discrete constants. In particular, it contains the generalized Sen black hole with mass, Newman-Unti-Tamburino parameter, charge, angular momentum, dilaton and axion parameters. So, to examine the WCCC for such a rich balck hole will be an important and interesting work. In this paper we will use the new version of the gedanken experiments to investigate the WCCC for this EMDA black hole.
	
	The paper is organized as follows. In Sect. \ref{ss2}, we review the Iyer-wald formalism to derive the variational identities. In Sect. \ref{ss3}, we focus on the EMDA theory of gravity and calculate the relevant quantities for our further analysis. In the end of this section the EMDA black hole solution are introduced. In Sect. \ref{ss4}, we present the set-up for the new version of the gedanken experiments, and derive the first order and second order perturbation inequalities for EMDA black holes. Whereafter, we investigate the new version of the gedanken experiments to destroy a near-extremal EMDA black hole. We prove the WCCC is valid under the second order correction of the perturbation. The conclusions are presented in Sect. \ref{ss5}.
	
	%###########################################################################%
	\section{Iyer-Wald Formalism and Varational Identities}\label{ss2}
	
	Firstly, we review the Iyer-Wald formalism to derive the variational identities \cite{Sorce:2017dst}. We consider a general diffeomorphism covariant theory of gravity on a 4-dimensional spacetime $\mathcal{M}$, where the Lagrangian 4-form $\mathbf{L}=\mathrm{L}\bm{\epsilon}$ is constructed locally by the metric field $g_{ab}$ and matter fields $\psi$'s, with $\bm{\epsilon}$ the volume element compatible with the metric $g_{ab}$. We denote the dynamic fields jointly as $\Phi=(g_{ab},\psi)$.
	The variation of $\mathbf{L}$ leads to
	\begin{equation}
	\delta \mathbf{L}(\Phi)=\mathbf{E}_{\Phi} \delta \Phi+\mathrm{d} \mathbf{\Theta}(\Phi, \delta \Phi) ,
	\end{equation}
	where $\mathbf{E}_{\Phi}=0$ gives the equations of motion (EOM), and $\mathbf{\Theta}$ is called the symplectic potential 3-form. The symplectic current 3-form is defined by
	\begin{equation}\label{fj2}
	\bm{\omega}(\Phi, \delta_1{\Phi}, \delta_2{\Phi})=\delta_1\mathbf{\Theta}(\Phi, \delta_2\Phi)-\delta_2\mathbf{\Theta}(\Phi, \delta_1\Phi) .
	\end{equation}
	The Noether current 3-form $\mathbf{J}_{\zeta}$ associated with a
	% arbitrary smooth
	vector field $\zeta$ is defined as
	\begin{equation}\label{fj3}
	\mathbf{J}_{\zeta}=\mathbf{\Theta}(\Phi, \mathscr{L}_{\zeta} \Phi)-\zeta \cdot \mathbf{L} ,
	\end{equation}
	where $\mathscr{L}_{\zeta}$ is the Lie derivative along the vector field $\zeta$. A straightforward calculation gives
	\begin{equation}
	\mathrm{d}\mathbf{J}_{\zeta}=-\mathbf{E}_{\Phi}\mathscr{L}_{\zeta} \Phi ,
	\end{equation}
	which implies $\mathrm{d}\mathbf{J}_{\zeta}=0$ when the EOM are satisfied. From Ref.~\citen{Iyer:1995kg}, the Noether current can also be expressed as
	\begin{equation}\label{fj5}
	\mathbf{J}_{\zeta}=\mathbf{C}_{\zeta}+\mathrm{d}\mathbf{Q}_{\zeta} ,
	\end{equation}
	where $\mathbf{Q}_{\zeta}$ is called the Noether charge and $\mathbf{C}_{\zeta}$ are the constraints of the theory, i.e., $\mathbf{C}_{\zeta}=0$ when the EOM are satisfied.
	Comparing the variations of Eqs.\eqref{fj3} and \eqref{fj5} with $\zeta$ fixed, we obtain the first variational identity
	\begin{equation}\label{fj6}
	\mathrm{d}\left[\delta\mathbf{Q}_{\zeta}-\zeta\cdot\mathbf{\Theta}(\Phi, \delta\Phi)\right]=\bm{\omega}(\Phi, \delta{\Phi}, \mathscr{L}_{\zeta}\Phi)-\zeta\cdot\mathbf{E} \, \delta\Phi-\delta\mathbf{C}_{\zeta} .
	\end{equation}
	The variation of \eqref{fj6} further gives the second variational identity
	\begin{equation}
	\mathrm{d}\left[\delta^2\mathbf{Q}_{\zeta}-\zeta\cdot\delta\mathbf{\Theta}(\Phi, \delta\Phi)\right]=\bm{\omega}(\Phi, \delta{\Phi}, \mathscr{L}_{\zeta}\delta{\Phi})-\zeta\cdot\delta\mathbf{E} \, \delta\Phi-\delta^2\mathbf{C}_{\zeta} .
	\end{equation}
	In what follows, we are interested in stationary axial-symmetric EMDA black hole solutions with horizon killing field
	\begin{equation}
	\xi^a=t^a+\Omega_\mathrm{H}\varphi^a ,
	\end{equation}
	where $t^a$, $\varphi^a$ and $\Omega_\mathrm{H}$ are the timelike killing field, axial killing field and the angular velocity of the horizon, respectively. For asymptotically flat black hole solutions, the perturbation of mass and angular momentum are given by
	\begin{align}
	\delta M=&\int_{\infty}\left[\delta\mathbf{Q}_t-t\cdot\mathbf{\Theta}(\Phi, \delta\Phi)\right] , \nonumber \\
	\delta J=&\int_{\infty}\delta\mathbf{Q}_{\varphi} .
	\end{align}
	We restrict consideration to the case where (a) $\Phi$ satisfies the EOM $\mathbf{E}_{\Phi}=0$, (b) $\xi^a$ is also a symmetry of the matter fields $\psi$, so that $\mathscr{L}_{\xi}\Phi=0$. Let $\Sigma$ be a hypersurface with a cross section $B$ on the horizon and with the spatial infinity as its boundaries. Using the Stokes theorem, we can obtain
	\begin{align}
	\delta M - \Omega_\mathrm{H} \delta J=& \int_B \left[ \delta \mathbf{Q}_{\xi} - \xi \cdot \mathbf{\Theta}(\Phi, \delta\Phi)  \right]-\int_{\Sigma} \delta \mathbf{C}_{\xi} , \label{fj10}  \\
	\delta^2 M - \Omega_\mathrm{H} \delta^2 J=&\int_B \left[ \delta^2 \mathbf{Q}_{\xi} - \xi \cdot \delta \mathbf{\Theta}(\Phi, \delta\Phi) \right]  \nonumber \\
	&-\int_{\Sigma} \xi \cdot \delta \mathbf{E} \, \delta \Phi-\int_{\Sigma} \delta^2 \mathbf{C}_{\xi}+\mathscr{E}_\Sigma(\Phi, \delta\Phi) , \label{fj11}
	\end{align}
	where the canonical energy is defined by
	\begin{equation}
	\mathscr{E}_\Sigma(\Phi, \delta\Phi)=\int_\Sigma \bm{\omega}(\Phi, \delta{\Phi}, \mathscr{L}_{\xi} \delta{\Phi}) .
	\end{equation}

	\section{Einstein-Maxwell-Dilaton-Axion Theory and Black Hole Solutions}\label{ss3}
	
	The EMDA black hole solution is given by the Lagrangian 4-form \cite{Garcia:1995qz}
	\begin{equation}
	\mathbf{L}=\frac{1}{16 \pi} \left[R-2(\partial \phi)^2-\frac{1}{2} e^{4 \phi} (\partial \mathrm{K})^2- e^{-2 \phi} F_{ab} F^{ab}-\mathrm{K} \, F_{ab} \star F^{ab} \right] \bm{\epsilon},
	\end{equation}
	which is obtained from a low-energy action of heterotic string theory, where $\star F_{ab}=\frac{1}{2} \epsilon_{abcd} F^{cd}$ is the dual of the electromagnetic field $F_{ab}$, $\phi$ is the massless dilaton field, and $\mathrm{K}$ is the axion field dual to the three-index antisymmetric tensor $\mathrm{H}=-\mathrm{exp}(4 \phi) \star \mathrm{d}\mathrm{K}/4$. In this paper we use the notation in Ref.~\citen{LozanoTellechea:1999my} and the $\epsilon$-tensor is defined as $\sqrt{-g} \epsilon^{0123}=-1$.
	
	The variation of the Lagrangian 4-form gives
	\begin{equation}
	\mathbf{E}_{\Phi} \delta \Phi=-\bm{\epsilon} \left[\frac{1}{2} T^{ab} \delta g_{ab}+j^a \delta A_a +E_\phi \delta \phi +E_{\mathrm{K}} \delta \mathrm{K} \right] ,
	\end{equation}
	where
	\begin{align}
	8\pi T^{ab}=& G^{ab}-8\pi T^{ab}_{EM}-8\pi T^{ab}_{DIL}-8\pi T^{ab}_{AX} , \nonumber \\
	j^a=& \frac{1}{4 \pi} \nabla_b \tilde F^{ab} ,  \nonumber \\
	E_\phi =& \frac{1}{16 \pi} \left[2 e^{4 \phi} (\partial \mathrm{K})^2-4\nabla_a \nabla^a \phi-2 e^{-2 \phi} F_{ab} F^{ab} \right] ,  \nonumber \\
	E_\mathrm{K}=& \frac{1}{16 \pi} \left[F_{ab} \star F^{ab}-\nabla_a (e^{4 \phi} \nabla^a \mathrm{K}) \right] ,
	\end{align}
	with
	\begin{align}
	G^{ab}=& R^{ab}-\frac{1}{2} R g^{ab} ,  \nonumber \\
	T^{ab}_{EM}=& \frac{1}{4 \pi} \left[ F^{ac} \tilde F^b{}_c-\frac{1}{4} g^{ab} F_{cd} \tilde F^{cd} \right] ,  \nonumber \\
	T^{ab}_{DIL}=& \frac{1}{8 \pi} \left[2 \nabla^a \phi \nabla^b \phi-(\partial \phi)^2 g^{ab} \right] ,  \nonumber \\
	T^{ab}_{AX}=& \frac{e^{4 \phi}}{32 \pi} \left[2 \nabla^a \mathrm{K} \nabla^b \mathrm{K}-(\partial \mathrm{K})^2 g^{ab} \right] ,
	\end{align}
	in which
	\begin{equation}
	\tilde F_{ab} \equiv e^{-2 \phi} F_{ab}+\mathrm{K} \star F_{ab}.
	\end{equation}
	Here $T^{ab}$ corresponds to the non-electromagnetic, -dilaton, and -axion (non-EDA) part of stress-energy tensor, and $j^a$ corresponds to the electromagnetic charge current.
	
	The symplectic potential 3-form is given by
	\begin{equation}
	\mathbf{\Theta}(\Phi, \delta \Phi)=\mathbf{\Theta}^{GR}(\Phi, \delta \Phi)+\mathbf{\Theta}^{EM}(\Phi, \delta \Phi)+\mathbf{\Theta}^{DIL}(\Phi, \delta \Phi)+\mathbf{\Theta}^{AX}(\Phi, \delta \Phi) ,
	\end{equation}
	where
	\begin{align}
	\mathbf{\Theta}_{abc}^{GR}(\Phi, \delta \Phi)=& \frac{1}{16 \pi} \epsilon_{dabc}  g^{de} g^{fg} (\nabla_g \delta g_{ef}-\nabla_{e} \delta g_{fg}) ,  \nonumber \\
	\mathbf{\Theta}_{abc}^{EM}(\Phi, \delta \Phi)=& -\frac{1}{4 \pi} \epsilon_{dabc} \tilde F^{de} \delta A_e ,  \nonumber \\
	\mathbf{\Theta}_{abc}^{DIL}(\Phi, \delta \Phi)=& -\frac{1}{4 \pi} \epsilon_{dabc} (\nabla^d \phi) \delta \phi ,  \nonumber \\
	\mathbf{\Theta}_{abc}^{AX}(\Phi, \delta \Phi)=& -\frac{1}{16 \pi} \epsilon_{dabc} (e^{4 \phi}\nabla^d \mathrm{K}) \delta \mathrm{K} .
	\end{align}
	From Eq.\eqref{fj2} we can obtain the symplectic current
	\begin{align}
	\bm{\omega}_{abc}(\Phi, \delta_1{\Phi}, \delta_2{\Phi})=& \bm{\omega}_{abc}^{GR}+\bm{\omega}_{abc}^{EM}+\bm{\omega}_{abc}^{DIL}+\bm{\omega}_{abc}^{AX} ,
	\end{align}
	where
	\begin{align}\label{fj21}
	\bm{\omega}_{abc}^{GR}=& \frac{1}{16 \pi} \epsilon_{dabc} w^d ,  \nonumber \\
	\bm{\omega}_{abc}^{EM}=& \frac{1}{4 \pi} \left[\delta_2 (\epsilon_{dabc} \tilde F^{de}) \delta_1 A_e-\delta_1 (\epsilon_{dabc} \tilde F^{de}) \delta_2 A_e \right] ,  \nonumber \\
	\bm{\omega}_{abc}^{DIL}=& \frac{1}{4 \pi} \left[\delta_2 (\epsilon_{dabc} \nabla^d \phi) \delta_1 \phi-\delta_1 (\epsilon_{dabc} \nabla^d \phi) \delta_2 \phi \right] ,  \nonumber \\
	\bm{\omega}_{abc}^{AX}=& \frac{1}{16 \pi} \left[\delta_2 (\epsilon_{dabc} e^{4 \phi} \nabla^d \mathrm{K}) \delta_1 \mathrm{K}-\delta_1 (\epsilon_{dabc} e^{4 \phi} \nabla^d \mathrm{K}) \delta_2 \mathrm{K} \right] ,
	\end{align}
	with
	\begin{align}
	{}& w^a=P^{abcdef} \left[\delta_2 g_{bc} \nabla_d \delta_1 g_{ef}-\delta_1 g_{bc} \nabla_d \delta_2 g_{ef} \right] ,  \nonumber \\
	{}& P^{abcdef}=g^{ae}g^{fb}g^{cd}-\frac{1}{2}g^{ad}g^{be}g^{fc}-\frac{1}{2}g^{ab}g^{cd}g^{ef}-\frac{1}{2}g^{bc}g^{ae}g^{fd}+\frac{1}{2}g^{bc}g^{ad}g^{ef} .
	\end{align}
	
	By using $\mathscr{L}_\xi g_{ab}=2\nabla_{ ( a } \xi_{b )}$, $\mathscr{L}_\xi A_a =\xi^b F_{ba}+\nabla_a (\xi^c A_c)$ and doing a straightforward calculation in Eq.\eqref{fj3}, we get the Noether current 3-form
	\begin{equation}
	\mathrm{J}_{abc}=\frac{1}{8 \pi} \epsilon_{dabc} \nabla_e \left[\nabla^{[e}\xi^{d]}+2\tilde{F}^{ed} A_f \xi^f \right]+\epsilon_{dabc} \left[T^d{}_f+A_f j^d \right] \xi^f .
	\end{equation}
	Comparing it with Eq.\eqref{fj5}, we obtain the Noether charge
	\begin{equation}
	(Q_\xi)_{ab}=(Q_\xi ^{GR})_{ab}+(Q_\xi ^{EM})_{ab} ,
	\end{equation}
	with
	\begin{align}\label{QEM}
	(Q_\xi ^{GR})_{ab}=&-\frac{1}{16 \pi} \epsilon_{abcd} \nabla^{[c}\xi^{d]} ,  \nonumber \\
	(Q_\xi ^{EM})_{ab}=&-\frac{1}{8 \pi} \epsilon_{abcd} \tilde{F}^{cd} A_e \xi^e ,
	\end{align}
	and the constraint term
	\begin{equation}
	(C_f)_{abc}=\epsilon_{dabc} \left[T^d{}_f+A_f j^d \right] .
	\end{equation}
	
	Now, we focus on the EMDA black hole solution given in Refs.~\citen{Garcia:1995qz,Jamil:2010vm}. The line element can be read off 
	\begin{align}\label{fj27}
	ds^2=&-\frac{\Xi-a^2 \, \mathrm{sin}^2\theta}{\Delta} dt^2-\frac{2a \, \mathrm{sin}^2\theta}{\Delta}\left[(r^2-2Dr+a^2)-\Xi \right]dt \, d\varphi  \nonumber \\
	&+\frac{\Delta}{\Xi}dr^2+\Delta d\theta^2+\frac{ \mathrm{sin}^2\theta}{\Delta}\left[(r^2-2Dr+a^2)^2-\Xi a^2 \, \mathrm{sin}^2\theta \right] d\varphi^2,
	\end{align}
	where
	\begin{align*}
	\Delta=& r^2-2Dr+a^2 \, \mathrm{cos}^2\theta, &\Xi& =r^2-2mr+a^2 ,  \\
	e^{2 \phi}=& \frac{\mathrm{W}}{\Delta}=\frac{\omega}{\Delta}(r^2+a^2 \, \mathrm{cos}^2\theta),  &\omega& =e^{2 \phi_0} ,  \\
	\mathrm{K}=& \mathrm{K}_0+\frac{2aD \, \mathrm{cos}\theta}{\mathrm{W}} ,
	&A_t&=\frac{1}{\Delta} (Qr-ga \, \mathrm{cos}\theta) ,  \\
	A_r=& A_\theta=0 ,
	&A_\varphi &=\frac{1}{a \Delta} (-Qra^2 \, \mathrm{sin}^2\theta+g(r^2+a^2)a \, \mathrm{cos}\theta) .
	\end{align*}
	The ADM mass $M$, angular momentum $J$, and dilaton charge $D$ are given by
	\begin{equation}
	M=m-D, \, \, \, \, \, \, \, \, J=a(m-D), \, \, \, \, \, \, \, \, D=-\frac{Q^2}{2 \omega M} ,
	\end{equation}
	respectively. The two horizons are
	\begin{equation}
	r_\pm=M+D \pm \sqrt{(M+D)^2-a^2},
	\end{equation}
	in which $r_+ $ is the event horizon. The surface gravity, area, angular velocity and electric potential of the horizon are given by
	\begin{align}
	\kappa=& \frac{r_+ -M-D}{r_+^2-2Dr_+ +a^2} ,  &\Omega_\mathrm{H}&=\frac{a}{r_+^2-2Dr_+ +a^2} ,  \nonumber \\
	\Phi_\mathrm{H}=& \frac{-2DM}{Q(r_+^2-2Dr_+ +a^2)} ,  &A_\mathrm{H}&=4 \pi (r_+^2-2Dr_+ +a^2).
	\end{align}
	The EMDA black hole becomes extremal when $(2 \omega M^2-Q^2)^2-4 \omega^2 J^2=0$. If
	\begin{equation}
	(2 \omega M^2-Q^2)^2-4 \omega^2 J^2 \ge 0 ,
	\end{equation}
	the metric describes a black hole solution, whereas the metric describes a naked singularity for the violation of the inequality.
	
	\section{Gedanken Experiments to Destroy a Near-Extremal EMDA Black Hole}\label{ss4}
	
	\subsection{Perturbation Inequalities of Gedanken Experiments}
	In this section, we use the new version of the gedanken experiment to obtain the first order and second order perturbation inequalities for the above EMDA black holes. We consider that the EMDA black holes are perturbed by a one-parameter family of the matter sources. The corresponding EOM can be expressed as
	\begin{align}
	G^{ab}(\lambda)=& 8 \pi \left[T^{ab}_{EM}(\lambda)+T^{ab}_{DIL}(\lambda)+T^{ab}_{AX}(\lambda)+T^{ab}(\lambda) \right] , \nonumber \\
	j^a(\lambda)=& \frac{1}{4 \pi} \nabla_b \tilde F^{ab}(\lambda) , \nonumber \\
	E_\phi (\lambda)=& 0, \, \, \, \, \, \, \, \, E_\mathrm{K} (\lambda)=0,
	\end{align}
	with $T^{ab}(0)=0$ and $j^a(0)=0$ for background spacetime. In this paper we consider the perturbation matter contains only the electromagnetic matter source, i.e., the sources of dilaton field $\phi$ and axion field $\mathrm{K}$ vanish, implied by $E_\phi (\lambda)=0$ and $E_\mathrm{K} (\lambda)=0$.
	
	As in Ref.~\citen{Sorce:2017dst}, we also assume the gedanken experiment admits the following assumptions.
	
	(a) All the perturbation matter goes into the black hole through a finite portion of the future horizon, i.e., the matter source $\delta T^{ab}$ and $\delta j^a$ are non-vanishing only in a compact region of future horizon.
	
	(b) Linear stability assumption. The non-extremal, unperturbed EMDA black hole is linearly stable to perturbation, i.e., any source free solution to the linearized EOM approaches a perturbation towards another EMDA black hole at sufficiently late times.
	
	(c) We choose a hypersurface $\Sigma=\mathrm{H} \cup \Sigma_1$ to perform our analysis. It starts from the bifurcate surface $B$ of the unperturbed horizon $\mathrm{H}$, continues up the future horizon through the matter source region of $\mathrm{H}$ till the very late cross section $B_1$ where the matter source vanishes, then becomes the spacelike hypersurface $\Sigma_1$ and continues out towards infinity. The boundaries of $\Sigma$ are located at the bifurcate surface and the spatial infinity.
	
	The linear stability assumption implies that the dynamic fields satisfy the source-free EOM, $\mathbf{E}[\Phi(\lambda)]=0$ on $\Sigma_1$, and the solutions are described by Eq.\eqref{fj27}.
	
	With the above set-up, we now derive the first order perturbation inequality at $\lambda=0$. Similar to the analysis in Ref.~\citen{Sorce:2017dst}, for a non-extremal black hole the horizon will be of bifurcate type, the second term of the first integral in Eq.\eqref{fj10} vanishes since $\xi^a=0$ on the bifurcate surface $B$. Therefore
	\begin{equation}\label{key}
	\int_B \left[ \delta \mathbf{Q}_{\xi} - \xi \cdot \mathbf{\Theta}(\Phi, \delta\Phi) \right]=\int_B  \delta \mathbf{Q}_{\xi} .
	\end{equation} 
	For the gravitational part, from the first expression of \eqref{QEM} we have
	\begin{equation}\label{key}
	\int_B  \delta \mathbf{Q}^{GR}_{\xi}=\frac{\kappa}{8\pi} \delta A_B,
	\end{equation}
	Where $A_B$ is the area of $B$ and $\kappa$ is the surface gravity of the event horizon. For the electromagnetic part, from the second expresson of \eqref{QEM} we obtain 
	\begin{equation}\label{key}
	\int_B  \delta \mathbf{Q}^{EM}_{\xi}=-\frac{1}{8 \pi}\int_B\left[ \xi^e A_e \delta (\epsilon_{abcd} \tilde{F}^{cd})+\xi^e (\delta A_e)\epsilon_{abcd} \tilde{F}^{cd}\right],
	\end{equation}
	where the second term vanishes at $B$ by $\xi^a|_B=0$, but $\xi^e A_e$ does not vanish since $\Phi_\mathrm{H}\equiv-\xi^e A_e (\lambda) $ must be constant on the horizon at $\lambda=0$. So,
	\begin{equation}\label{key}
	\int_B  \delta \mathbf{Q}^{EM}_{\xi}=\frac{1}{8 \pi} \Phi_\mathrm{H}\int_B \delta (\epsilon_{abcd} \tilde{F}^{cd})=\Phi_\mathrm{H} \delta Q_B,
	\end{equation} 
	Where $Q_B$ is the electric charge flux integral over $B$. 
	The assumption that the perturbation vanishes on the bifurcate surface $B$ leads to $\delta A_B=\delta Q_B=0$. This also holds for extremal black holes. Therefore, the first integral vanishes in Eq.\eqref{fj10}.  By using the fact that $T^{ab}=j^a=0$ in the background spacetime (since $\mathbf{E}[\Phi(0)]=0$), then the Eq.\eqref{fj10} can be written as
	\begin{align}\label{fj33}
	\delta M - \Omega_\mathrm{H} \delta J=& -\int_{\Sigma} \delta \mathbf{C}_{\xi} \nonumber \\
	=&- \int_\mathrm{H} \epsilon_{dabc} \delta T^d{}_e \xi^e-\int_\mathrm{H} A_e \xi^e \epsilon_{dabc} \delta j^d  \nonumber \\
	=& \int_\mathrm{H} \tilde{\epsilon}_{abc} \delta T_{de} k^d \xi^e +\Phi_\mathrm{H} \delta Q  \nonumber \\
	\ge & \Phi_\mathrm{H} \delta Q,
	\end{align}
	where $\tilde{\epsilon}_{abc}$ is the volume element on $\mathrm{H}$, which is defined by $\epsilon_{dabc}=-4k_{[d}\tilde{\epsilon}_{abc]}$ with the future-directed normal vector $k^a \propto \xi^a$, and $\int_\mathrm{H} \epsilon_{dabc} \delta j^d=\delta Q_{flux}=\delta Q$ is the total flux of charge through the horizon. In the last line we used the null energy condition $\delta T_{ab}k^a k^b|_\mathrm{H} \ge 0$ for the non-EDA stress-energy tensor.
	Thus, we obtain the first order perturbation inequality
	\begin{equation}\label{fj34}
	\delta M - \Omega_\mathrm{H} \delta J - \Phi_\mathrm{H} \delta Q \ge 0 .
	\end{equation}
	Under the first order perturbation, if we want to violate $(2 \omega M^2-Q^2)^2-4 \omega^2 J^2 \ge 0$, the optimal choice is to saturate \eqref{fj34} by requiring $\delta T_{ab}k^a k^b|_\mathrm{H} = 0$, i.e., the energy flux through the horizon vanishes for the first order non-EDA perturbation. Then, \eqref{fj33} reduces to
	\begin{equation}\label{fj35}
	\delta M - \Omega_\mathrm{H} \delta J - \Phi_\mathrm{H} \delta Q =0.
	\end{equation}
	
	Next, we derive the second order perturbation inequality. In exact parallel to the derivation of the first order inequality \eqref{fj33}, Eq.\eqref{fj11} becomes
	\begin{equation}\label{sieq1}
	\delta^2 M - \Omega_\mathrm{H} \delta^2 J=-\int_{\mathrm{H}} \xi \cdot \delta \mathbf{E} \, \delta \Phi-\int_{\mathrm{H}} \delta^2 \mathbf{C}_{\xi}+\mathscr{E}_\Sigma(\Phi, \delta\Phi) ,
	\end{equation}
	with
	\begin{align}
	(\xi \cdot \delta \mathbf{E} \, \delta \Phi)_{abc}=& -\xi^d \epsilon_{dabc} \left[\frac{1}{2} \delta T^{ef} \delta g_{ef}+\delta j^e \delta A_e \right] ,  \nonumber \\
	(\delta^2 \mathbf{C}_\xi)_{abc}=& \delta^2 (\epsilon_{dabc} T^d{}_e \xi^e)+\delta^2 (\epsilon_{dabc} A_e \xi^e j^d) .
	\end{align}
	In Eq.(\ref{sieq1}), the integrals in the first two terms only depend on the surface $\mathrm{H}$ since $ \delta \mathbf{E}=\delta^2 \mathbf{C}_\xi=0 $ on $\Sigma_1$ by the assumption that there are no sources outside the black hole at late times. In addition, since $\xi^a$ is tangent to the horizon, the first term of the right side of (\ref{sieq1}) vanishes. By using the condition $\xi^a \delta A_a =0 $ on $\mathrm{H}$ from a gauge transformation, Eq.(\ref{sieq1}) becomes
	\begin{align}\label{fj38}
	\delta^2 M - \Omega_\mathrm{H} \delta^2 J=& \mathscr{E}_\Sigma(\Phi, \delta  \Phi)+\int_{\mathrm{H}} \tilde{\epsilon}_{abc} \xi^e k^d \delta^2 T_{de} + \Phi_\mathrm{H} \delta^2 Q  \nonumber \\
	\ge & \mathscr{E}_{\Sigma_1}(\Phi, \delta \Phi)+\mathscr{E}_\mathrm{H}(\Phi, \delta \Phi)+ \Phi_\mathrm{H} \delta^2 Q ,
	\end{align}
	where we have used the optimal choice $\delta T_{ab}k^a k^b|_\mathrm{H} = 0$ and the null energy condition $\delta^2 T_{ab}k^a k^b|_\mathrm{H} \ge 0$ for the second order perturbed non-EDA stress-energy tensor. To obtain $\mathscr{E}_\mathrm{H}(\Phi, \delta \Phi)$, we split it into
	\begin{equation}
	\mathscr{E}_\mathrm{H}(\Phi, \delta \Phi)=\int_\mathrm{H} \bm{\omega}^{GR}+\int_\mathrm{H} \bm{\omega}^{EM}+\int_\mathrm{H} \bm{\omega}^{DIL}+\int_\mathrm{H} \bm{\omega}^{AX} .
	\end{equation}
	The contribution of the gravitational part has already been calculated in Ref.~\citen{Sorce:2017dst}
	\begin{equation}
	\int_\mathrm{H} \bm{\omega}^{GR}=\frac{1}{4 \pi} \int_\mathrm{H} \tilde{\bm{\epsilon}} \, (\xi^a \nabla_a u) \delta \sigma_{bc}\delta \sigma^{bc} \ge 0 .
	\end{equation}
	For the electromagnetic, dilaton and axion parts, according to \eqref{fj21}, the symplectic currents $\bm{\omega}(\Phi, \delta{\Phi}, \mathscr{L}_{\xi} \delta{\Phi})$ are given as
	\begin{align}
	\bm{\omega}_{abc}^{EM}=& \frac{1}{4 \pi} \epsilon_{dabc} \left[ \mathscr{L}_\xi \delta \tilde{F}^{de} \delta A_e - \delta \tilde{F}^{de} \mathscr{L}_\xi \delta A_e \right] \nonumber \\
	& +\frac{1}{4 \pi} \left[(\mathscr{L}_\xi \delta \epsilon_{dabc}) \tilde{F}^{de} \delta A_e - \delta \epsilon_{dabc} \tilde{F}^{de} \mathscr{L}_\xi \delta A_e \right] ,  \nonumber \\
	\bm{\omega}_{abc}^{DIL}=& \frac{1}{4 \pi} \epsilon_{dabc} \left[ \mathscr{L}_\xi \delta (\nabla^d \phi ) \delta \phi-\delta (\nabla^d \phi ) \mathscr{L}_\xi \delta \phi \right] \nonumber \\
	& +\frac{1}{4 \pi} \left[  \mathscr{L}_\xi \delta \epsilon_{dabc} (\nabla^d \phi ) \delta \phi - \delta \epsilon_{dabc} (\nabla^d \phi ) \mathscr{L}_\xi \delta \phi \right] ,  \nonumber \\
	\bm{\omega}_{abc}^{AX}=& \frac{1}{16 \pi} \epsilon_{dabc} \left[ \mathscr{L}_\xi \delta (e^{4 \phi} \nabla^d \mathrm{K} ) \delta \mathrm{K}-\delta (e^{4 \phi} \nabla^d \mathrm{K} ) \mathscr{L}_\xi \delta \mathrm{K} \right] \nonumber \\
	& +\frac{1}{16 \pi} \left[  \mathscr{L}_\xi \delta \epsilon_{dabc} (e^{4 \phi} \nabla^d \mathrm{K} ) \delta \mathrm{K} - \delta \epsilon_{dabc} (e^{4 \phi} \nabla^d \mathrm{K} ) \mathscr{L}_\xi \delta \mathrm{K} \right] .
	\end{align}
	Through the similar calculations to those in Refs.~\citen{Sorce:2017dst} and \citen{Jiang:2019ige}, the corresponding contributions to the canonical energy give the following inequalities
	\begin{align}\label{fj42}
	\int_\mathrm{H} \bm{\omega}^{DIL}=& \int_\mathrm{H} \tilde{\bm{\epsilon}} \, \xi^a k^b \delta^2 T_{ab}^{DIL} \ge 0 , \nonumber \\
	\int_\mathrm{H} \bm{\omega}^{AX}=& \int_\mathrm{H} \tilde{\bm{\epsilon}} \, \xi^a k^b \delta^2 T_{ab}^{AX} \ge 0 ,  \nonumber \\
	\int_\mathrm{H} \bm{\omega}^{EM}=& \frac{1}{2 \pi} \int_\mathrm{H} \tilde{\bm{\epsilon}} \,  k_d \xi^f \delta \tilde{F}^{de} \delta F_{ef}
	=\int_\mathrm{H} \tilde{\bm{\epsilon}} \, \xi^a k^b \delta^2 T_{ab}^{EM} \ge 0 .
	\end{align}
	Again the null energy conditions for the second order perturbed matter fields stress-energy tensors have been used in \eqref{fj42}. Then, \eqref{fj38} reduces to
	\begin{equation}
	\delta^2 M - \Omega_\mathrm{H} \delta^2 J - \Phi_\mathrm{H} \delta^2 Q \ge \mathscr{E}_{\Sigma_1}(\Phi, \delta \Phi) .
	\end{equation}
	What remains now is to calculate $\mathscr{E}_{\Sigma_1}(\Phi, \delta \Phi)$. We adopt the trick in Ref.~\citen{Sorce:2017dst}, and write $\mathscr{E}_{\Sigma_1}(\Phi, \delta \Phi)=\mathscr{E}_{\Sigma_1}(\Phi, \delta \Phi^F)$, where $\Phi^F(\alpha)$ denotes a field configuration of another one-parameter family of EMDA black hole solutions with parameters given by
	\begin{align}
	M^F(\alpha)=& M+\alpha \delta M ,  \nonumber \\
	J^F(\alpha)=& J+\alpha \delta J ,  \nonumber \\
	Q^F(\alpha)=& Q+\alpha \delta Q ,
	\end{align}
	where $\delta M$, $\delta J$ and $\delta Q$ are chosen to agree with the corresponding values for our first order perturbation of $\Phi(\lambda)$. Then, for this family we have $ \delta^2 M = \delta^2 J = \delta^2 Q_B=\delta \mathbf{E}=\delta^2 \mathbf{C}_\xi =\mathscr{E}_\mathrm{H}(\Phi, \delta \Phi^F)=0$. According to Eq.\eqref{fj11} we have
	\begin{equation}
	\mathscr{E}_{\Sigma_1}(\Phi, \delta \Phi^F)=-\int_B \left[ \delta^2 \mathbf{Q}_{\xi} - \xi \cdot \delta \mathbf{\Theta}(\Phi, \delta\Phi^F) \right]=-\frac{\kappa}{8 \pi}\delta^2 A_B^F ,
	\end{equation}
	thus we obtain the second order perturbation inequality
	\begin{equation}\label{fj46}
	\delta^2 M - \Omega_\mathrm{H} \delta^2 J - \Phi_\mathrm{H} \delta^2 Q \ge -\frac{\kappa}{8 \pi}\delta^2 A_B^F .
	\end{equation}
	Taking two variations of the area formula $A_B=4 \pi (r_+^2-2Dr_+ +a^2)$, we obtain
	\begin{align}
	\delta^2 A_B^F=& -\frac{\pi}{\omega^4 M^6 \epsilon^3} \big[2 \omega \big(Q^6-2 \omega M^2 Q^4 (3+\epsilon)-4 \omega^2 Q^2(J^2-M^4(3+2\epsilon))  \nonumber \\
	& -8 \omega^3 M^2 (M^4(1+\epsilon)-J^2(3+\epsilon)) \big) (\delta M)^2-\big(Q^6-2 \omega M^2 Q^4 (3+\epsilon)  \nonumber \\
	& +4 \omega^2 Q^2(-3 J^2+M^4(3+2\epsilon))+8 \omega^3 M^2 (J^2-M^4)(1+\epsilon) \big) (\delta Q)^2  \nonumber \\
	& +2 \omega^2 (2 \omega M^2 -Q^2)^2(\delta J)^2+8 \omega^2 J Q (2 \omega M^2 -Q^2)\delta  Q\delta J  \nonumber \\
	& -32 \omega^3 M Q J^2\delta M \delta Q+16 \omega^3 M J  (2 \omega M^2 -Q^2)\delta M \delta J \big] ,
	\end{align}
	in which
	\begin{equation}
	\epsilon=\frac{\sqrt{(2 \omega M^2 -Q^2)^2-4\omega^2 J^2}}{2 \omega M^2} .
	\end{equation}
	For the near-extremal black hole, $\epsilon$ is a small parameter, then the surface gravity of EMDA black hole can be expressed as
	\begin{equation}
	\kappa= \frac{\omega M \epsilon}{2 \omega M^2(1+\epsilon)-Q^2}.
	\end{equation}
	Expanding the right side of \eqref{fj46} to lowest order in $\epsilon$, we obtain
	\begin{equation}\label{fj50}
	\delta^2 M - \Omega_\mathrm{H} \delta^2 J - \Phi_\mathrm{H} \delta^2 Q \ge -\frac{Q^2 (\delta Q)^2}{4 \omega^3 M^5 \epsilon^2} \big(Q^2+2 \omega M-2 \omega M^2 \big)^2 ,
	\end{equation}
	where we have used the Eq.\eqref{fj35} to eliminate $\delta M$ from the expression.

	\subsection{Near-Extremal EMDA Black Hole Cannot be Over-Charged or Over-Spun
	}
	
	With the above preparation, we now investigate the new version of the gedanken experiments to over-charge or over-spin a near-extremal EMDA black hole. We consider a one-parameter family $\Phi (\lambda)$, and the background spacetime $\Phi (0)$ is a near-extremal EMDA black hole, $\epsilon \ll 1$. Define a function of $\lambda$ as
	\begin{equation}
	h(\lambda)=\left[2 \omega M(\lambda)^2-Q(\lambda)^2 \right]^2-4 \omega^2 J(\lambda)^2 .
	\end{equation}
	The WCCC is violated if there exists a solution $\Phi (\lambda)$ such that $h(\lambda)<0$. Expanding $h(\lambda)$ to second order in $\lambda$, we have
	\begin{align}\label{fj52}
	h(\lambda)&= (2 \omega M^2-Q^2)^2-4 \omega^2 J^2  \nonumber \\
	& +\lambda\big[8 \omega M (2 \omega M^2-Q^2) \delta M-4 Q (2 \omega M^2-Q^2) \delta Q-8 \omega^2 J \delta J \big] \nonumber \\
	& +\frac{1}{2} \lambda^2 \big[(8 \omega(2 \omega M^2-Q^2)+32 \omega^2 M^2) (\delta M)^2+8 \omega M(2 \omega M^2-Q^2) \delta^2 M  \nonumber \\
	& -8 \omega^2 (\delta J)^2-8 \omega^2 J \delta^2 J -32 \omega M Q \delta M \delta Q-4Q(2 \omega M^2-Q^2)\delta^2 Q  \nonumber \\
	&+(8Q^2-4(2 \omega M^2-Q^2)) (\delta Q)^2 \big]+O(\lambda^3) .
	\end{align}
	If we consider only the linear term of $\lambda$ in \eqref{fj52}, then by using the inequality \eqref{fj34}, $h(\lambda)$ is constrained by
	\begin{equation}
	h(\lambda) \ge 4\omega^2 M^4 \epsilon^2+4(Q^2+2 \omega M-2 \omega M^2) Q \delta Q \lambda +O(\lambda^2) ,
	\end{equation}
	which implies that it is possible to make $h(\lambda)<0$, i.e., the black hole could be over-charged or over-spun.
	
	Next, we include the $O(\lambda^2)$ term in \eqref{fj52}. By using inequality \eqref{fj50} and for optimal choice \eqref{fj35}, we have
	\begin{align}
	h(\lambda) \ge & 4\omega^2 M^4 \epsilon^2+4(Q^2+2 \omega M-2 \omega M^2) Q \delta Q \lambda \nonumber \\
	& +\frac{(Q^2+2 \omega M-2 \omega M^2)^2 Q^2 (\delta Q)^2}{\omega^2 M^4 \epsilon^2} \lambda^2+O(\lambda^3)  \nonumber \\
	=& \left[2 \omega M^2 \epsilon+\frac{(Q^2+2 \omega M-2 \omega M^2) Q \delta Q}{ \omega M^2 \epsilon} \lambda \right]^2+O(\lambda^3) .
	\end{align}
	Thus, no violation of $(2 \omega M^2-Q^2)^2-4 \omega^2 J^2 \ge 0$ can occur when the second order correction of the perturbation is taken into account, i.e., this near-extremal EMDA black hole cannot be over-charged or over-spun. In our case the parameter $ \omega $ can be any nonzero number, which characterizes the black hole hairs (Dilaton, Axion, etc.). So, it is implied that the validity of WCCC is not relevant to the black hole hairs.
	
	Note that, the EMDA solutions contain the Kerr-Sen black hole \cite{Sen:1992ua} as a special case, which can be seen from the metric \eqref{fj27} when $\omega=1$ and $D=-m \, sinh^2(\alpha/2)$. In this case, our result reduced to that obtained in Ref.~\citen{Jiang:2019vww} that the near-extremal Kerr-Sen black hole cannot be over-charged or over-spun on the level of the second order approximation.
	
	\section{Conclusions and remarks}\label{ss5}
	In this paper, we have used the new version of the gedanken experiments to examine the WCCC for an EMDA black hole. We derived the first order and second order inequalities relating the mass, angular momentum and electric charge in this framework. We show that no violations of WCCC can occur with the increase of the background solution parameters for a near-extremal EMDA black hole when the second order correction of the perturbation was taken into account. The result implies that once an EMDA black hole is formed, it will never be destroyed by being over-charged or over-spun. When the parameters $\omega=1$ and $D=-m \, sinh^2(\alpha/2)$ are taken, the EMDA metric becomes the Kerr-Sen one, and our conclusion reduces to that in Ref.~\citen{Jiang:2019vww} that the WCCC is preserved for a Kerr-Sen black hole.
	
	For EMDA black hole we have shown that the validity of WCCC is not affected by the increase of the background solution parameters when the perturbation matter contains only the electromagnetic matter source. In Ref.~\citen{Jiang:2020btc}, it is shown that the WCCC is restored in the Einstein-Maxwell gravity with scalar hairs when the scalar perturbation is considered but the background conserved scalar charge \cite{Pacilio:2018gom} is not included in the perturbation inequalities. It is accessible that a complete analysis of WCCC in the new version of the gedanken experiments may contain both the background conserved scalar charge in the perturbation inequalities and the perturbation matter with scalar field when the black holes have some scalar hairs. This will be an important issue to study the WCCC in the future work.

\end{document}